\documentclass[twocolumn, %reprint,
 amsmath,amssymb,
prb,
floatfix,
]{revtex4}
\usepackage{graphicx}% Include figure files
\usepackage{dcolumn}% Align table columns on decimal point
\usepackage{bm}% bold math
\usepackage{epstopdf}
\usepackage{color}
\begin{document}

%\preprint{APS/123-QED}

\title{Effect of uniaxial strain on the site occupancy of hydrogen in vanadium from density-functional calculations}% Force line breaks with \\

\author{Robert Johansson}
\email{robert.johansson@physics.uu.se}
\author{Rajeev Ahuja}
\affiliation{Department of Physics \& Astronomy, Uppsala University, Box 516, SE-751 20 Uppsala, Sweden}
\author{Olle Eriksson}
\affiliation{Department of Physics \& Astronomy, Uppsala University, Box 516, SE-751 20 Uppsala, Sweden}
\author{Bj\"{o}rgvin Hj\"{o}rvarsson}
\affiliation{Department of Physics \& Astronomy, Uppsala University, Box 516, SE-751 20 Uppsala, Sweden}
\author{Ralph H. Scheicher}
 \email{ralph.scheicher@physics.uu.se}
\affiliation{Department of Physics \& Astronomy, Uppsala University, Box 516, SE-751 20 Uppsala, Sweden}

\date{\today}% It is always \today, today,

\begin{abstract}
We investigate the influence of uniaxial strain on site occupancy of hydrogen vanadium, using density functional theory. The site occupancy is found to be strongly influenced by the strain state of the lattice. The results provide the conceptual framework of the atomistic description of the observed hysteresis in the $\alpha$ to $\beta$ phase transition in bulk, as well as the preferred octahedral occupancy of hydrogen in strained V layers.
%\begin{description}

%71.15.Mb, 88.30.rd, 68.60.Bs
%\end{description}
\end{abstract}

%Furthermore, the lattice expansion is determined by the concentration and site occupancy of hydrogen which can result in a self-amplified change of site occupancy.

%\pacs{71.15.Mb,88.30.rd,62.20.-x,68.55.Ln,61.72.Dd,68.65.Cd}% PACS, the Physics and Astronomy
\maketitle

%\tableofcontents

\section{Introduction}
Vanadium is a transition metal with electronic configuration $[Ar]3d^{3}4s^{2}$ that forms a body-centered cubic structure (bcc). The bcc structure contains tetrahedral and octahedral interstitial sites that can accommodate hydrogen \cite{alefeld1,alefeld2}. In bulk vanadium the hydrogen is found to reside in tetrahedral sites at low concentrations ($\alpha$-phase), while in the $\beta$-phase H occupies octahedral sites\cite{alefeld1,alefeld2}.
The symmetry of the hydrogen induced local strain field is strongly depending on the site occupancy, which is reflected in the hydrogen induced expansion of the lattice\cite{alefeld1,alefeld2}. When hydrogen resides in a tetrahedral site, the local strain field is close to spherical, while it is almost uniaxial when hydrogen resides in the octahedral sites.

\begin{figure}[ht]
\begin{center}
\includegraphics[angle=0,width=0.45\textwidth]{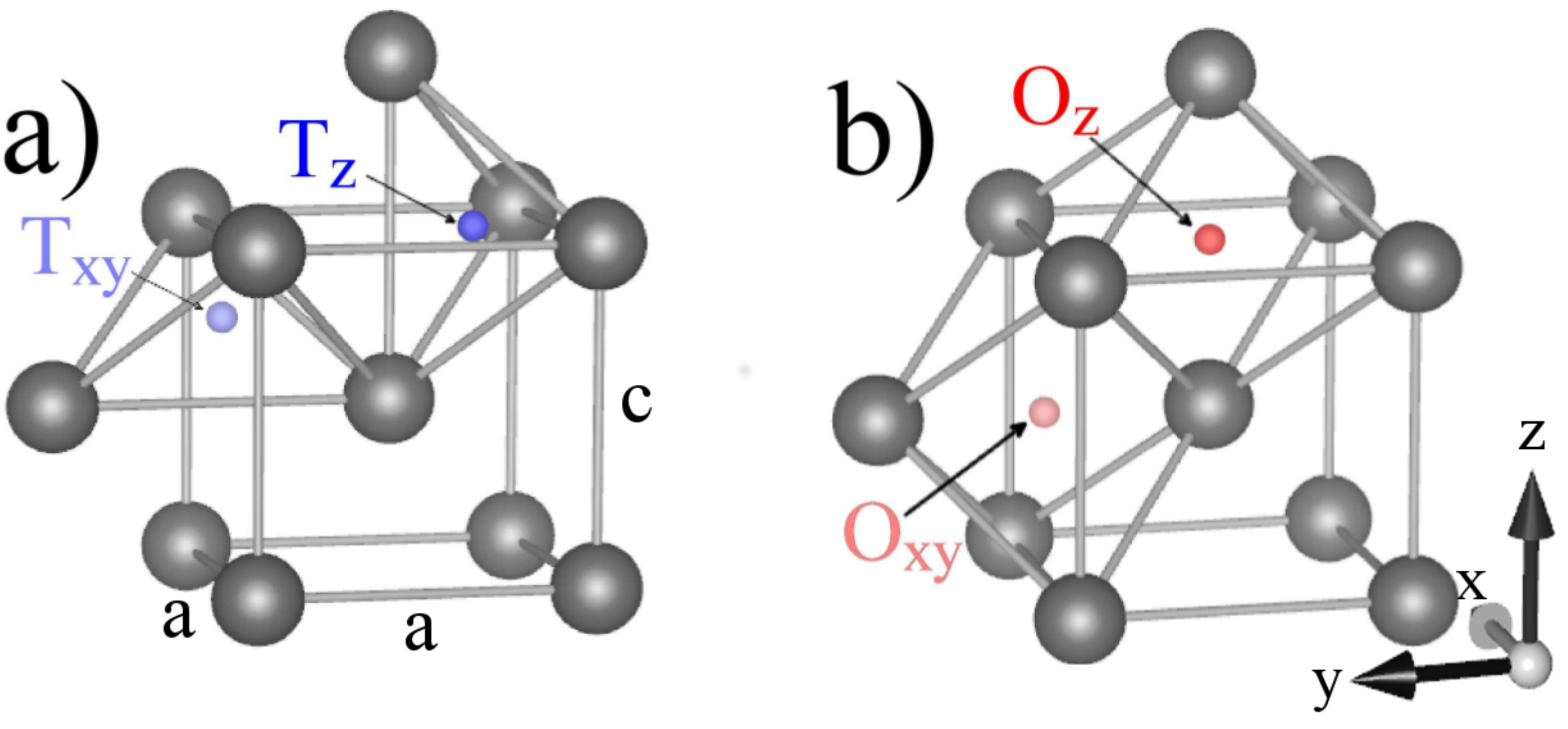}
\caption{(Color online) The different types of (a) tetrahedral and (b) octahedral interstitial sites in bcc vanadium are illustrated here. Large dark spheres represent vanadium atoms and the small red, blue, light blue, and light red spheres represent (according to their respective labels) different interstitial positions that hydrogen can occupy. The $z$-axis is aligned along the vertical direction, while the $x$- and $y$-axes lie in the horizontal plane.}
\label{OT}
\end{center}
\end{figure}

In a body-centered tetragonal structure there are three types of tetrahedral sites and three types of octahedral sites and there are in total three octahedral and six tetrahedral sites per metal atom (Figure \ref{OT}). The six tetrahedral sites comprise of four $T_z$ and two $T_{xy}$ sites ($T_{xy}$ refers to either one of the equivalent $T_x$ or $T_y$ sites). The three octahedral sites comprise of one $O_z$ and two $O_{xy}$ sites.
The local elastic response of the lattice arising from the presence of hydrogen in these interstitial sites \cite{buck,alefeld4,pundt1,pundt2,pundt3, anden} gives rise to a local strain field, which is the cause of the hydrogen induced expansion\cite{alefeld4}. The expansion can be viewed as the sum of the hydrogen induced local strain fields and can therefore, in principle, be used to determine the preferred site occupancy of hydrogen.

The volume changes depend strongly on the boundary conditions and site occupancy, enabling the polarization of the local strain fields. This can, for example, be experimentally accomplished by the use of clamping \cite{palsson12, alefeld4} as the preferred site occupancy of hydrogen in vanadium is linked to the strain state of the structure \cite{EXAFS,HSL,Hinfev,olsson}.
The hydrogen induced volume changes of a clamped epitaxial film is restricted to the direction perpendicular to the surface.
For example, a single crystal vanadium (001) film on a MgO(001) substrate will exhibit a lattice expansion (or contraction) in the (001) direction, independent of site occupancy \cite{EXAFS, palsson12}. Clamping of V can therefore be used to change the volume expansion, as only 1/3 of the strain field will propagate to the surface, when hydrogen is residing in tetrahedral sites. By the same token, the uniaxial component of the local strain field arising from a $O_z$ site occupancy will reach the free boundaries and therefore not restrict the expansion.  Furthermore,  when hydrogen resides in $O_x$ or $O_y$ sites, the local strain field can not give rise to any expansion due to the constraint implemented through elastic boundary conditions (clamping). Clamping and straining a V layer will therefore strongly affect the site occupancy, and thereby the observed volume changes \cite{palsson12,EXAFS,olsson,alefeld4}.

The goal of the present computational study is to provide a conceptual framework on the influence of strain on site occupancy in Vanadium. The results are generalised and can be used as a guidance when discussing hydrogen occupancy in any transition metal. Thus, we use computational methods to perform "virtual experiments" using an atomistic framework and first-principles methodology to explore the influence of polarization of the hydrogen induced local strain fields generated in bulk vanadium.  The results provide a plausible atomistic description of the $O_z$ site occupancy at low concentrations in strained V layers\cite{palsson12}, change of site occupancy with hydrogen concentration in biaxially strained lattices  as well as providing insight on the origin of the observed hysteresis \cite{alefeld2} in hydrogen absorption and desorption in bulk V.\\

\section{Methods}
The calculations were performed using the Vienna \textit{Ab initio} Simulation Package (VASP) \cite{kressevasp1,kressevasp2,kressevasp3,kressevasp4}. The interactions between the electrons and the nuclei was obtained using the projector-augmented-wave method \cite{kresse,bloechl}. The Perdew-Burke-Ernzerhof (PBE) \cite{perdew,perdew2} approach was employed to approximate the exchange and correlation terms in the density functional theory (DFT) \cite{DFT,KS} method. A conjugate gradient algorithm was used to relax the atomic nuclei positions to a local minimum in the total energy landscape.

In order to reduce H-H interactions resulting from the imposed periodic boundary conditions while still studying a system that is small enough to be computationally manageable for a large number of calculations, a supercell consisting of 128 vanadium atoms ($4\times 4\times 4$ bcc unit cells) was constructed to mimic bulk vanadium in which the lowest possible ratio of hydrogen to vanadium [H/V] is 1/128 (corresponding to 0.775 at.\% of hydrogen). Due to the rather large dimensions (11.9 {\AA} $\times$ 11.9 {\AA} $\times$ 11.9 {\AA}) of the supercell, only the $\Gamma$ point was used in sampling the Brillouin zone. Comparisons of the total energy, using a 3 $\times$ 3 $\times$ 3 k-point mesh further established that $\Gamma$ point sampling of the Brillouin zone is sufficient for a quantitative investigation. Zero-point energy calculations were deliberately excluded, not because they were unimportant, but rather because their correct treatment requires and deserves a separate dedicated study.

Calculations for higher hydrogen concentrations than [H/V] = 1/128 were carried out by randomly distributing hydrogen atoms into the vanadium supercell and calculating the average volume of 50 structures, with different hydrogen distributions. Hydrogen concentrations [H/V] of 8/128 (5.88 at.\%), 16/128 (11.11 at.\%), 32/128 (20.00 at.\%) and 64/128 (33.33 at.\%) were investigated. This corresponds to a disordered state, mimicking the conditions above the phase boundaries of a $\beta$-phase.

The $x$ and $y$ lattice vectors were fixed for all calculations, only allowing lattice relaxation in the $z$-direction. The motivation for this approach is to mimic the experimental conditions of hydrogen uptake in a superlattice where the bottom layer of a thin film of vanadium is held in place through strong bonds to a substrate. This constraint results in a one dimensional lattice expansion, perpendicular to the substrate's plane. This approach also mimics the bct lattice of the $\beta$ phase when $c/a$ $\simeq$ 1.1 (see Figure \ref{OT}).

It is sufficient to consider the strain in one direction, for example the $z$-direction, to capture the effect of strain on site occupancy. This implies also that we can treat the $O_x$ and $O_y$ sites to be equivalent when the hydrogen is occupying $O_z$ sites in the $\beta$-phase. These are identical in the sense that a rotation by 90$^\circ$ around the $z$-axis will map the $O_x$ site onto the $O_y$ site and vice versa.

\section{Results \& discussion}

\subsection{Strain and site occupancy}
Before presenting the results from \textit{ab initio} calculations we will provide a conceptual framework for the effect of strain on site occupancy, using a simple hard-sphere model. We will use this approach to estimate the relative energetics of hydrogen occupation in octahedral and tetrahedral sites, solely based on the available interstitial space in a V single crystal. The maximum sphere radius that can be accommodated in the interstitial space formed by metal atom spheres arranged in a bcc lattice is 0.155 for octahedral and 0.291 for tetrahedral sites, in units of the metal atom sphere radius. In the atomistic model used here the vanadium has a Wigner-Seitz radius of 1.217~{\AA} and the corresponding value for hydrogen is 0.370~{\AA}. An octahedral site in V has therefore $0.155\cdot1.217 $\AA$=0.189$~{\AA} of spherical radius available, which is much smaller than the hydrogen radius. A tetrahedral site provides $0.291\cdot1.217 $\AA$=0.354$~{\AA}, which is close to the hydrogen radius. From this consideration, one can see directly that it is not energetically favourable for hydrogen to occupy octahedral sites in an unstrained lattice, because of a large overlap between H and V electrons, raising the total energy through a Born-Mayer repulsion \cite{born-mayer}. In the tetrahedral sites, the corresponding density overlap is much smaller, favouring occupation of tetrahedral sites. When the lattice is under uniaxial tensile strain (i.e., $c/a > 1$) the maximum sphere radius that can be accommodated in the $T_z$ and $O_z$ sites is more and more shifted in favour of the $O_z$ sites. The maximum sphere radius that can be accommodated in $T_z$ and $O_z$ sites becomes equal for $c/a = 1.118$.

When the lattice is expanded in the $z$-direction the $T_z$ and $O_z$ sites are energetically favored in comparison to their $x$ and $y$ oriented counterparts. This is rather obvious for the octahedral sites but not as clear for the tetrahedral sites. For the $O_z$ sites, a tensile strain in the $z$-direction will increase the spacing between the vanadium atoms that sit closest to hydrogen, while for $O_{xy}$ sites, the closest vanadium atoms lie in the $xy$-plane, which are geometrically unaffected by the uniaxial strain in the $z$-direction. In the experiments by P\'{a}lsson \textit{et al.} \cite{palsson12} no distinction is made between hydrogen occupation of $T_z$ and $T_{xy}$ sites.

\begin{figure}[ht]
\begin{center}
\includegraphics[angle=0,width=0.45\textwidth]{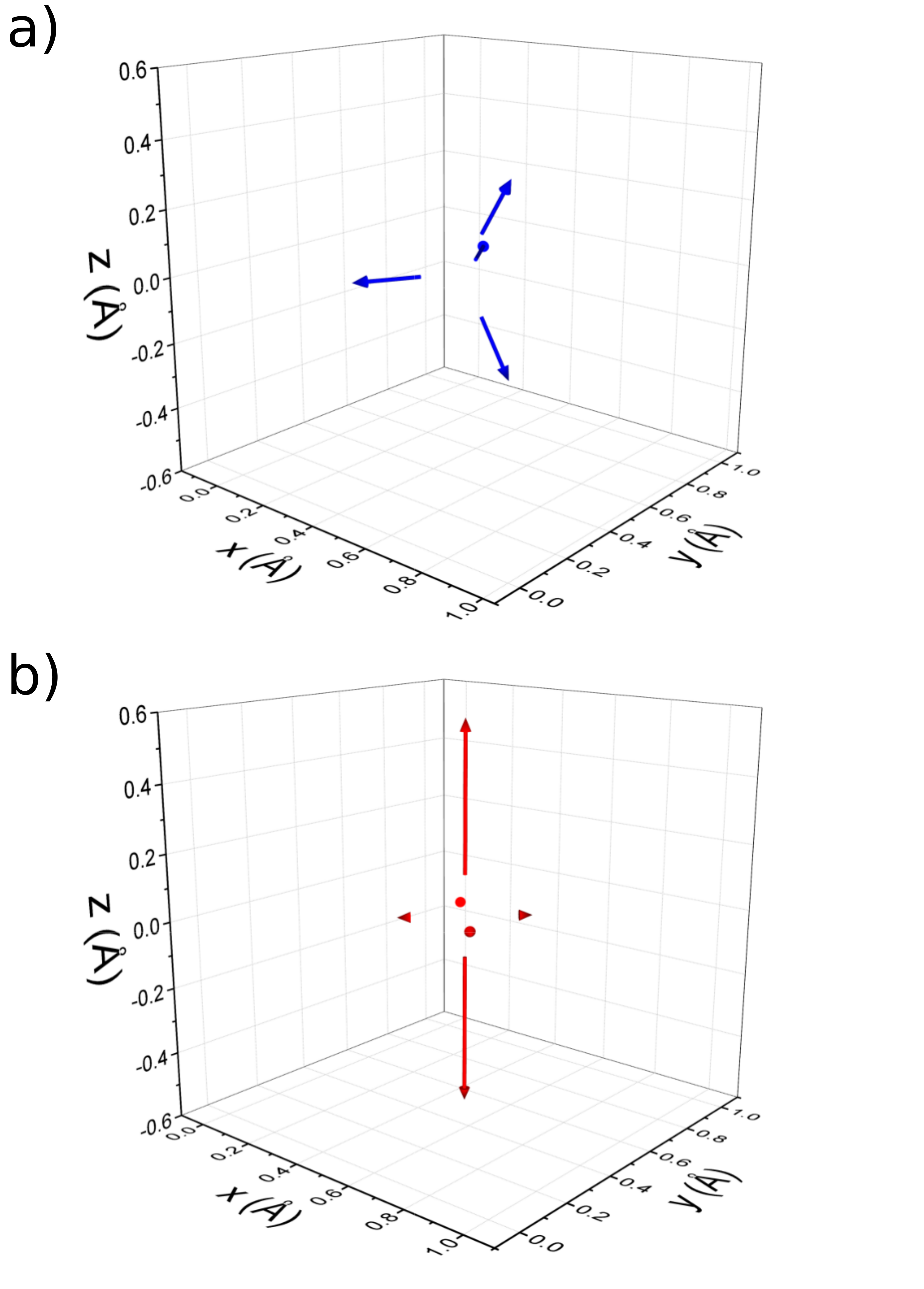}
\caption{(Color online) The strain on surrounding vanadium atoms by hydrogen occupying a (a) $T_z$ site or (b) $O_z$ site. The arrows represent the displacement vectors, i.e., how much the V atoms are ``pushed'' away by the H atom. The length of the arrows has been scaled by a factor of 30.}
\label{strain}
\end{center}
\end{figure}

%, since the difference in available interstitial space to fill for occupation in $T_z$ and $T_{xy}$ sites is almost negligible
%Up until this point, the discussion is based solely on geometrical considerations alone.
Figure \ref{strain} shows the \textit{ab initio} calculated local strain fields in vanadium caused by hydrogen occupying either a $T_z$ or an $O_z$ site in a 128 vanadium atoms supercell. The arrows indicate the direction and the magnitude of the displacement of the vanadium atoms. Only the strain on the vanadium atoms in the $O_z$ and $T_z$ sites are shown (i.e., 4 atoms for a tetrahedral site and 6 atoms for an octahedral site).  The isotropic strain field from hydrogen occupying a tetrahedral site and the strongly anisotropic strain field from occupying an octahedral site is clearly seen in Figure \ref{strain}.  The ``top'' and ``bottom'' vanadium atoms in the octahedron (i.e., the two vanadium atoms that possess the same $x$ and $y$ coordinates) are much closer to the hydrogen than the vanadium atoms in the tetrahedron; hence, the former are pushed farther away. In the absence of hydrogen the calculated lattice parameter is 2.99 {\AA} .
%This outcome is expected based  from symmetry considerations illustrated in Figure \ref{OT}.
%Without the influence of a local strain field from hydrogen occupying an $O_z$ site and with no amount of hydrogen in the system, the distance between the ``top'' and ``bottom'' vanadium atoms equals that of the calculated lattice constant of 2.99 {\AA} for vanadium.
When hydrogen is placed in the $O_z$ site, the ``top'' and ``bottom'' V-atoms are displaced, increasing their mutual distance to 3.35 {\AA}. This local strain corresponds to an increase of 12.1\% in spacing between the V atoms which is in excellent agreement with the experimental results of 12.7\% for $\beta$-phase VH$_{0.5}$ obtained by EXAFS \cite{EXAFS}.

\begin{figure}[ht]
\begin{center}
\includegraphics[angle=0,width=0.45\textwidth]{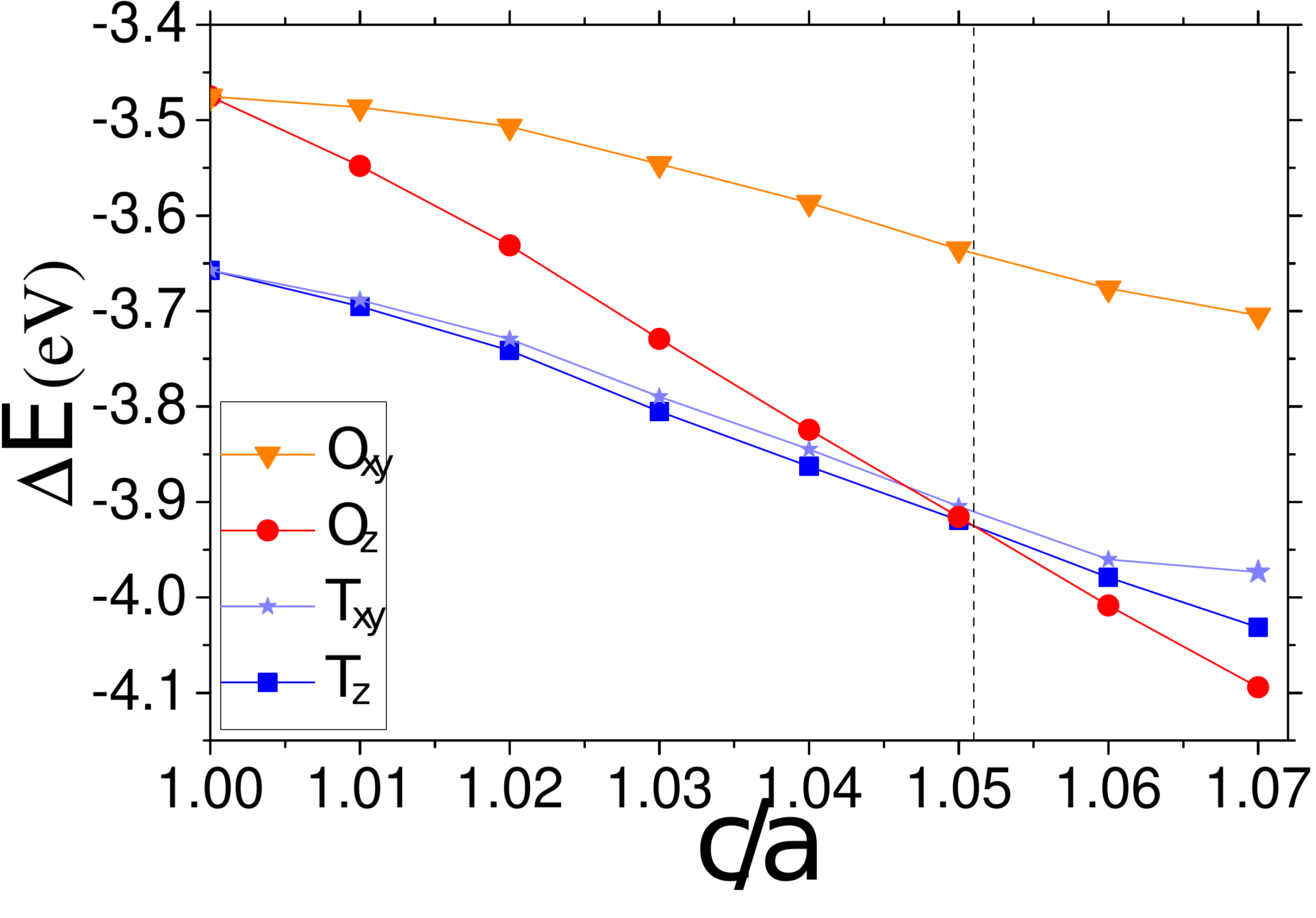}
\caption{(Color online) Energy difference as a function of an externally applied global uniaxial lattice strain $c/a$ where $\Delta E = E(V+H)-E(V)$. The dashed vertical line at $c/a = 1.051$ marks the critical uniaxial lattice strain for which hydrogen occupancy of $T_z$ and $O_z$ sites becomes energetically equivalent.}
\label{energy}
\end{center}
\end{figure}

To quantify the qualitative ideas obtained from the hard-sphere model, we used \textit{ab initio} total energy calculations to determine the preferred hydrogen occupancy.
Figure \ref{energy} shows a plot of the energy for a single hydrogen in a supercell occupying a $T_z$, $T_{xy}$, $O_z$ or an $O_{xy}$ site as a function of the $c/a$ ratio ([H/V] = 1/128, corresponding to 0.775 at.\% of hydrogen). The vertical axis shows $\left[E(V+H)-E(V)\right]$ where $E(V+H)$ is the total energy of the metal-hydrogen system and $E(V)$ is the total energy of the hydrogen free vanadium supercell, both calculated at the same $c/a$ ratio. When the lattice is uniaxially strained ( $c/a > 1$), the $O_z$ sites will ``open up'' as described above and  becomes more energetically favoured. This is easily inferred from the results in Figure \ref{energy} since the slope of the $O_z$ line is larger than than that of $T_z$, thus at some  $c/a$ ratio the site occupancy of $O_z$ will be favoured as compared to the T sites. As seen in Figure \ref{strain}, the strain is very large in the $z$-direction for hydrogen occupying an $O_z$ site. A comparison of the strain fields from hydrogen occupation of $O_z$ and $T_z$ sites shows a larger increase of available spherical radius for the hydrogen occupying an $O_z$ site. This, together with the favourable effect of the uniaxial tensile strain for the occupation of the $O_z$ sites makes the $O_z$ sites energetically favorable already when $c/a > 1.051$. The hard sphere model yielded a transition at $c/a = 1.118$ which can be considered as satisfying when considering the simplicity of the model.

%from the hard-sphere model is due to the circumstance that in this model, the local strain fields were not taken into account. Furthermore, any effects from the electronic structure and directional binding are not considered either in the hard-sphere model, but do naturally enter in the first-principles calculations.
%strength of an externally applied global uniaxial strain field, quantified by the
% when it comes to increasing the available spherical radius for hydrogen occupation,

\subsection{Concentration dependence of site occupancy}
\begin{figure}[ht]
\begin{center}
\includegraphics[angle=0,width=0.45\textwidth]{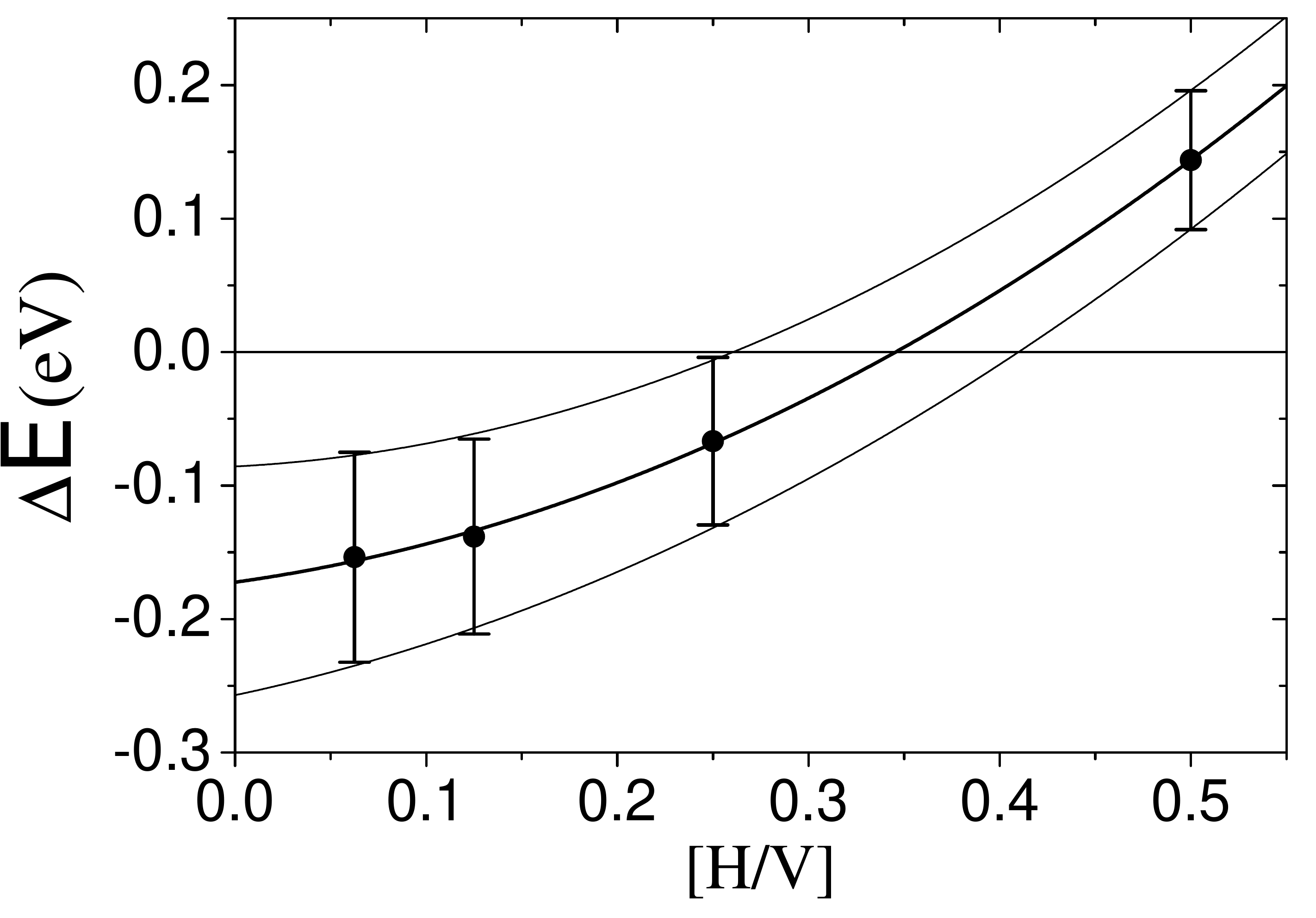}
\caption{Energy difference between $T_z$ and $O_z$ occupancy at optimal $c/a$ ratios as a function of hydrogen concentration. $\Delta E = (E_{T_z}-E_{O_z})/N_H$ where $N_H$ is the number of hydrogen atoms in the simulation. Data points are for average values and the bars indicate $\pm$ the standard deviation (defined as the square root of the variance). Lines are second order polynomial fits.}
\label{evsconc}
\end{center}
\end{figure}

It is not only the initial strain state which is the source of tetragonal distortion. The hydrogen induced volume changes will also influence the $c/a$ ratio in clamped samples and thereby alter the energy balance between the $T_z$ and $O_z$ sites.
Figure \ref{evsconc} compares the energies of $T_z$ and $O_z$ site occupancy at optimal $c/a$ ratios to identify the critical hydrogen concentration where change in site occupancy occurs. The average energy of 50 structures with random hydrogen distributions for four different hydrogen concentrations is calculated (disordered phase). Change in site occupancy is approximated to occur between [H/V] of 0.34$\pm$0.07 as this is where the total energy of $T_z$ and $O_z$ site occupancy becomes equal.

The shift in the site occupancy is found to be driven by energetics rather than entropy. The configurational entropy was determined using Boltzmann's entropy formula and the internal energy was approximated as the number of hydrogen atoms in $O$-sites times an energy penalty of 0.2 eV  (i.e., we approximate that moving a hydrogen atom from a $T$ site to an $O$ site will raise the energy by 0.2 eV, in accordance with the difference in energy between $T$ and $O$ site occupancy, cf. Figure 3). For all tested hydrogen concentrations and for a broad temperature range, the internal energy is always found to dominate the entropy part, so that coexistence of $T$ and $O$ sites is concluded unlikely to occur in the low concentration region. The hydrogen induced lattice expansion can however give rise to change in site occupancy. This can take place in both ordered and disordered phases, thus a $O_z$ occupancy does not need to imply an ordered $\beta$-phase and a change of site does therefore not by necessity imply a disorder-ordered phase transition.

%The combinatorics of a quantitative investigation is too complicated and beyond the scope of this study.

\subsection{Volume expansion and hysteresis effects}
\begin{figure}[ht]
\begin{center}
\includegraphics[angle=0,width=0.45\textwidth]{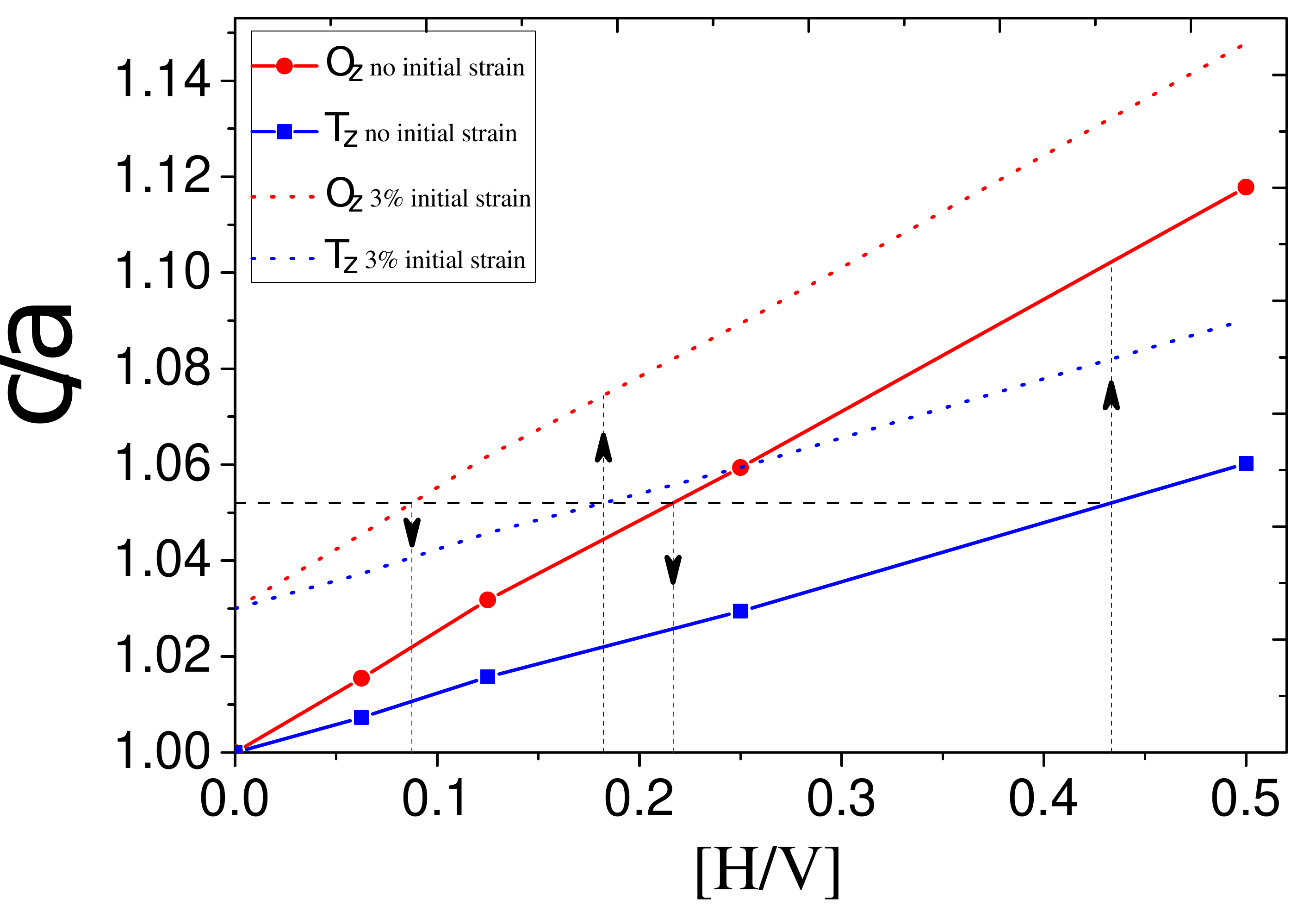}
\caption{(Color online) The uniaxial lattice strain $c/a$ resulting from varying the concentration of hydrogen occupying exclusively either $T_z$ sites (blue data points and lines) or $O_z$ sites (red data points and lines) in bcc vanadium. The horizontal black dashed line at $c/a = 1.051$ marks the critical uniaxial lattice strain for which the hydrogen occupancy of $T_z$ and $O_z$ site becomes equal in energy, as seen in Figure \ref{energy}. The vertical colored full lines indicate at which hydrogen concentration the critical $c/a$ ratio of 1.051 is reached for occupancy of $T_z$ ([H/V] = 0.424) and $O_z$ ([H/V] = 0.212) sites, respectively, when there is no initial strain (i.e. $c/a = 1.00$). The dotted lines represent the case of an initial strain of $c/a = 1.03$ before any hydrogen has entered the system. The critical $c/a$ ratio is then reached at [H/V] = 0.173 for $T_z$ occupancy and 0.084 for $O_z$ occupancy.}
\label{volume}
\end{center}
\end{figure}

Figure \ref{volume} shows the resulting uniaxial lattice expansion (quantified by the $c/a$ ratio) as a function of hydrogen concentration [H/V]. The relationship between calculated $c/a$ ratio and hydrogen concentration [H/V] is to a very good approximation linear with a slope of 0.120 for $T_z$ sites and 0.236 for $O_z$ sites, see Figure \ref{volume}. These results are in excellent agreement with the experimental results by P\'{a}lsson \textit{et al.} \cite{palsson12} which determined the expansion to be 0.1189(7) for hydrogen occupation in tetrahedral sites and 0.19(1) for octahedral sites.  The calculated change in total volume due to hydrogen occupation of $T_z$ and $O_z$ sites are 1.61 \AA$^3$ and 3.14 \AA$^3$, respectively per added hydrogen atom.
%As shown above the $T_z$ or $O_z$ sites results in profound difference in the lattice expansion. %Two of the strain components are clamped ($x$ and $y$), restricted by the substrate in the experiments and by boundary conditions in the calculations. %The $x$ and $y$ lattice vectors have been kept fixed, only allowing the lattice expansion in the $z$-direction (clamping).
$O_z$ occupancy gives rise to larger increase in volume, as compared to $T_z$ occupancy, due to the anisotropy of the local strain field  as seen in Figure \ref{strain}. The strain component in the $z$-direction is larger for $O_z$ than for $T_z$ sites, implying that a shift from $T_z$ to $O_z$ occupancy is accompanied by an increased $c/a$ ratio, which favours $O_z$ occupancy.  The shift in site occupancy from $T_z$ to $O_z$ can therefore be viewed as a self-amplified process. The shift in site occupancy resembles therefore in many ways a first order phase transition.

 %This results in an increase in the $c/a$ ratio will cause the $O_z$ sites to become energetically more favourable as compared to $T_z$ sites.
%Now we will discuss the influence of hydrogen loading and unloading, upon the obtained site occupancy.
Now we will discuss the difference in the lattice response when the hydrogen concentration is increased or decreased.
In an unstrained or nearly unstrained lattice hydrogen is exclusively found in $T_z$ sites . When increasing the hydrogen concentration from low concentrations, the expansion will open up the $O_z$ sites, which become energetically favored above the critical $(c/a)_{crit}$ value of 1.051. The uniaxial lattice expansion will therefore result in a shift in site occupancy from $T_z$ to $O_z$ at that concentration. $(c/a)_{crit}$ for the change from $T_z$ to $O_z$ occupancy is marked by a vertical line in Figure \ref{energy} and a horizontal dashed line in Figure \ref{volume}.

When starting at a high concentration all the hydrogen will reside in $O_z$ sites. When decreasing the concentration, $(c/a)_{crit}$ =1.051 will be reached at [H/V] = 0.212, resulting in a shift in site occupancy from $O_z$ to $T_z$. Thus, when increasing the concentration the shift from $T_z$ to $O_z$ is reached at a different concentration as compared to the change of sites from $O_z$ to $T_z$ sites when decreasing the hydrogen concentration. Thus, a hysteresis with respect to lattice expansion is expected when loading and unloading H under the specified conditions and the thermal excitations are smaller than the energy difference between the two states. These effects do resemble the $\alpha$ to $\beta$ phase transition in bulk V, with respect to both change of sites as well as observed hysteresis  \cite{alefeld2}. Furthermore, these results clearly illustrate the effect of clamping on the site occupancy, which can be changed without entering the $\beta$-phase in V. When the initial strain of the sample is changed, these boundaries will move as illustrated in Figure \ref{volume}: With a biaxial compressive strain in the $x-y$ plane, the boundaries will move to lower concentrations and the hysteresis gap will decrease. When $c/a$ will be larger than a critical value, hydrogen will solely reside in  $O_z$ sites, as inferred from experiments\cite{palsson12}.\\
\\
\\
\\
\\
\\

\section{Summary}
The preferred interstitial site occupancy in vanadium with constrained boundaries has been studied using calculations based on density functional theory. The energetics of hydrogen atoms in a bcc-bct supercell were investigated to provide a conceptual understanding of the experimentally observed shifts in site occupancy \cite{palsson12}. In the investigated range of $c/a$ from 1.00 up to 1.07, the tetrahedral ($T_z$) sites are energetically favored for hydrogen occupation in comparison to the octahedral ($O_z$) sites in the $c/a$ range from 1.00 to 1.051. The octahedral sites are energetically favored for hydrogen occupation when  $c/a$ $>$ 1.051. The forces exerted on the vanadium lattice by hydrogen atoms occupying interstitial sites will alter the global strain state which in turn triggers a shift in site occupancy above the critical value of $c/a$ $=$ 1.051. This self-amplified process can be understood by the obtained strain field from octahedral ($O_z$) site occupancy which has a larger $z$-component than that obtained for a tetrahedral ($T_z$) site occupancy. The increase in $c/a$ as a function of hydrogen concentration [H/V] is linear and in good agreement with previously obtained experimental results.

The different slopes of $c/a$ as a function of [H/V] for tetrahedral ($T_z$) and octahedral ($O_z$) site occupancy has the consequence that the condition for shift in site occupancy is met at different hydrogen concentrations [H/V] when starting from high ($O_z$) or low concentration ($T$). This leads to the theoretical prediction of a hysteresis in the hydrogen loading-unloading process, in which the switch from $T_z$ to $O_z$ site occupancy and the reverse switch from $O_z$ to $T_z$ occur at different  hydrogen concentrations. The experimentally observed \cite{palsson12} coexistence of tetrahedral and octahedral hydrogen occupation in the [H/V]  concentration range of 0.065--0.068 might be an indication that such a hysteresis behavior could indeed be found in vanadium.

%--1.070 and are expected to be so also beyond that $c/a$ ratio since the maximum sphere radius which can be accommodated in the interstitial space will continue to keep favoring the octahedral ($O_z$) sites.

\section{Acknowledgments}
Financial support from the Swedish Research Council is gratefully acknowledged. The project is part of the COST Action MP1103. O.E. also acknowledges the KAW foundation, eSSENCE, STandUP for Energy, and the ERC (project 247062 - ASD). The calculations were performed on resources provided by the Swedish National Infrastructure for Computing (SNIC).


\begin{thebibliography}{24}
\expandafter\ifx\csname natexlab\endcsname\relax\def\natexlab#1{#1}\fi
\expandafter\ifx\csname bibnamefont\endcsname\relax
  \def\bibnamefont#1{#1}\fi
\expandafter\ifx\csname bibfnamefont\endcsname\relax
  \def\bibfnamefont#1{#1}\fi
\expandafter\ifx\csname citenamefont\endcsname\relax
  \def\citenamefont#1{#1}\fi
\expandafter\ifx\csname url\endcsname\relax
  \def\url#1{\texttt{#1}}\fi
\expandafter\ifx\csname urlprefix\endcsname\relax\def\urlprefix{URL }\fi
\providecommand{\bibinfo}[2]{#2}
\providecommand{\eprint}[2][]{\url{#2}}

\bibitem[{\citenamefont{Alefeld and V{\"o}lkl}(1978{\natexlab{a}})}]{alefeld1}
\bibinfo{author}{\bibfnamefont{G.}~\bibnamefont{Alefeld}} \bibnamefont{and}
  \bibinfo{author}{\bibfnamefont{J.}~\bibnamefont{V{\"o}lkl}},
  \emph{\bibinfo{title}{Hydrogen in Metals I}}, vol.~\bibinfo{volume}{28}
  (\bibinfo{publisher}{Springer}, \bibinfo{year}{1978}{\natexlab{a}}).

\bibitem[{\citenamefont{Alefeld and V{\"o}lkl}(1978{\natexlab{b}})}]{alefeld2}
\bibinfo{author}{\bibfnamefont{G.}~\bibnamefont{Alefeld}} \bibnamefont{and}
  \bibinfo{author}{\bibfnamefont{J.}~\bibnamefont{V{\"o}lkl}},
  \emph{\bibinfo{title}{Hydrogen in Metals II}}, vol.~\bibinfo{volume}{29}
  (\bibinfo{publisher}{Springer}, \bibinfo{year}{1978}{\natexlab{b}}).

\bibitem[{\citenamefont{P\'alsson et~al.}(2012)\citenamefont{P\'alsson,
  W\"alde, Amft, Wu, Ahlberg, Wolff, Pundt, and Hj\"orvarsson}}]{palsson12}
\bibinfo{author}{\bibfnamefont{G.~K.} \bibnamefont{P\'alsson}},
  \bibinfo{author}{\bibfnamefont{M.}~\bibnamefont{W\"alde}},
  \bibinfo{author}{\bibfnamefont{M.}~\bibnamefont{Amft}},
  \bibinfo{author}{\bibfnamefont{Y.}~\bibnamefont{Wu}},
  \bibinfo{author}{\bibfnamefont{M.}~\bibnamefont{Ahlberg}},
  \bibinfo{author}{\bibfnamefont{M.}~\bibnamefont{Wolff}},
  \bibinfo{author}{\bibfnamefont{A.}~\bibnamefont{Pundt}}, \bibnamefont{and}
  \bibinfo{author}{\bibfnamefont{B.}~\bibnamefont{Hj\"orvarsson}},
  \bibinfo{journal}{Phys. Rev. B} \textbf{\bibinfo{volume}{85}},
  \bibinfo{pages}{195407} (\bibinfo{year}{2012}).

\bibitem[{\citenamefont{Burkert et~al.}(2001)\citenamefont{Burkert, Miniotas,
  and Hj{\"o}rvarsson}}]{EXAFS}
\bibinfo{author}{\bibfnamefont{T.}~\bibnamefont{Burkert}},
  \bibinfo{author}{\bibfnamefont{A.}~\bibnamefont{Miniotas}}, \bibnamefont{and}
  \bibinfo{author}{\bibfnamefont{B.}~\bibnamefont{Hj{\"o}rvarsson}},
  \bibinfo{journal}{Physical Review B} \textbf{\bibinfo{volume}{63}},
  \bibinfo{pages}{125424} (\bibinfo{year}{2001}).

\bibitem[{\citenamefont{Hj\"{o}rvarsson
  et~al.}(1997)\citenamefont{Hj\"{o}rvarsson, Andersson, and Karlsson}}]{HSL}
\bibinfo{author}{\bibfnamefont{B.}~\bibnamefont{Hj\"{o}rvarsson}},
  \bibinfo{author}{\bibfnamefont{G.}~\bibnamefont{Andersson}},
  \bibnamefont{and} \bibinfo{author}{\bibfnamefont{E.}~\bibnamefont{Karlsson}},
  \bibinfo{journal}{J. Alloys Compd.} \textbf{\bibinfo{volume}{253}},
  \bibinfo{pages}{51} (\bibinfo{year}{1997}).

\bibitem[{\citenamefont{Andersson et~al.}(1999)\citenamefont{Andersson,
  Andersson, and Hj{\"o}rvarsson}}]{Hinfev}
\bibinfo{author}{\bibfnamefont{G.}~\bibnamefont{Andersson}},
  \bibinfo{author}{\bibfnamefont{P.~H.} \bibnamefont{Andersson}},
  \bibnamefont{and}
  \bibinfo{author}{\bibfnamefont{B.}~\bibnamefont{Hj{\"o}rvarsson}},
  \bibinfo{journal}{Journal of Physics} \textbf{\bibinfo{volume}{11}},
  \bibinfo{pages}{6669} (\bibinfo{year}{1999}).

\bibitem[{\citenamefont{Olsson and Hj\"orvarsson}(2005)}]{olsson}
\bibinfo{author}{\bibfnamefont{S.}~\bibnamefont{Olsson}} \bibnamefont{and}
  \bibinfo{author}{\bibfnamefont{B.}~\bibnamefont{Hj\"orvarsson}},
  \bibinfo{journal}{Phys. Rev. B} \textbf{\bibinfo{volume}{71}},
  \bibinfo{pages}{035414} (\bibinfo{year}{2005}).

\bibitem[{\citenamefont{Buck and Alefeld}(1972)}]{buck}
\bibinfo{author}{\bibfnamefont{H.}~\bibnamefont{Buck}} \bibnamefont{and}
  \bibinfo{author}{\bibfnamefont{G.}~\bibnamefont{Alefeld}},
  \bibinfo{journal}{Phys. Stat. Sol} \textbf{\bibinfo{volume}{49}},
  \bibinfo{pages}{317} (\bibinfo{year}{1972}).

\bibitem[{\citenamefont{Alefeld}(1972)}]{alefeld4}
\bibinfo{author}{\bibfnamefont{G.}~\bibnamefont{Alefeld}},
  \bibinfo{journal}{Ber. Bunsenges. Phys. Chem} \textbf{\bibinfo{volume}{76}},
  \bibinfo{pages}{746} (\bibinfo{year}{1972}).

\bibitem[{\citenamefont{Laudahn
  et~al.}(1999{\natexlab{a}})\citenamefont{Laudahn, F{\"a}hler, Krebs, and
  Pundt}}]{pundt1}
\bibinfo{author}{\bibfnamefont{U.}~\bibnamefont{Laudahn}},
  \bibinfo{author}{\bibfnamefont{S.}~\bibnamefont{F{\"a}hler}},
  \bibinfo{author}{\bibfnamefont{H.~U.} \bibnamefont{Krebs}}, \bibnamefont{and}
  \bibinfo{author}{\bibfnamefont{A.}~\bibnamefont{Pundt}},
  \bibinfo{journal}{Appl. Phys. Lett.} \textbf{\bibinfo{volume}{74}},
  \bibinfo{pages}{647} (\bibinfo{year}{1999}{\natexlab{a}}).

\bibitem[{\citenamefont{Dornheim et~al.}(2003)\citenamefont{Dornheim, Pundt,
  Kirchheim, v.~d. Molen, Kooij, Kerssemakers, Griessen, Harms, and
  Geyer}}]{pundt2}
\bibinfo{author}{\bibfnamefont{M.}~\bibnamefont{Dornheim}},
  \bibinfo{author}{\bibfnamefont{A.}~\bibnamefont{Pundt}},
  \bibinfo{author}{\bibfnamefont{R.}~\bibnamefont{Kirchheim}},
  \bibinfo{author}{\bibfnamefont{S.~J.} \bibnamefont{v.~d. Molen}},
  \bibinfo{author}{\bibfnamefont{E.~S.} \bibnamefont{Kooij}},
  \bibinfo{author}{\bibfnamefont{J.}~\bibnamefont{Kerssemakers}},
  \bibinfo{author}{\bibfnamefont{R.}~\bibnamefont{Griessen}},
  \bibinfo{author}{\bibfnamefont{H.}~\bibnamefont{Harms}}, \bibnamefont{and}
  \bibinfo{author}{\bibfnamefont{U.}~\bibnamefont{Geyer}},
  \bibinfo{journal}{Journal of Applied Physics} \textbf{\bibinfo{volume}{93}},
  \bibinfo{pages}{8958} (\bibinfo{year}{2003}).

\bibitem[{\citenamefont{Laudahn
  et~al.}(1999{\natexlab{b}})\citenamefont{Laudahn, Pundt, Bicker, von
  H{\"u}lsen, Geyer, Wagner, and Kirchheim}}]{pundt3}
\bibinfo{author}{\bibfnamefont{U.}~\bibnamefont{Laudahn}},
  \bibinfo{author}{\bibfnamefont{A.}~\bibnamefont{Pundt}},
  \bibinfo{author}{\bibfnamefont{M.}~\bibnamefont{Bicker}},
  \bibinfo{author}{\bibfnamefont{U.}~\bibnamefont{von H{\"u}lsen}},
  \bibinfo{author}{\bibfnamefont{U.}~\bibnamefont{Geyer}},
  \bibinfo{author}{\bibfnamefont{T.}~\bibnamefont{Wagner}}, \bibnamefont{and}
  \bibinfo{author}{\bibfnamefont{R.}~\bibnamefont{Kirchheim}},
  \bibinfo{journal}{Journal of Alloys and Compounds}
  \textbf{\bibinfo{volume}{293–-295}}, \bibinfo{pages}{490 }
  (\bibinfo{year}{1999}{\natexlab{b}}).

\bibitem[{\citenamefont{Blomqvist et~al.}(2010)\citenamefont{Blomqvist,
  P\'alsson, Ara\'ujo, Ahuja, and Hj{\"o}rvarsson}}]{anden}
\bibinfo{author}{\bibfnamefont{A.}~\bibnamefont{Blomqvist}},
  \bibinfo{author}{\bibfnamefont{G.~K.} \bibnamefont{P\'alsson}},
  \bibinfo{author}{\bibfnamefont{C.~M.} \bibnamefont{Ara\'ujo}},
  \bibinfo{author}{\bibfnamefont{R.}~\bibnamefont{Ahuja}}, \bibnamefont{and}
  \bibinfo{author}{\bibfnamefont{B.}~\bibnamefont{Hj{\"o}rvarsson}},
  \bibinfo{journal}{Phys. Rev. Lett.} \textbf{\bibinfo{volume}{105}},
  \bibinfo{pages}{185901} (\bibinfo{year}{2010}).

\bibitem[{\citenamefont{Kresse and Hafner}(1993)}]{kressevasp1}
\bibinfo{author}{\bibfnamefont{G.}~\bibnamefont{Kresse}} \bibnamefont{and}
  \bibinfo{author}{\bibfnamefont{J.}~\bibnamefont{Hafner}},
  \bibinfo{journal}{Phys. Rev. B} \textbf{\bibinfo{volume}{47}},
  \bibinfo{pages}{558} (\bibinfo{year}{1993}).

\bibitem[{\citenamefont{Kresse and Hafner}(1994)}]{kressevasp2}
\bibinfo{author}{\bibfnamefont{G.}~\bibnamefont{Kresse}} \bibnamefont{and}
  \bibinfo{author}{\bibfnamefont{J.}~\bibnamefont{Hafner}},
  \bibinfo{journal}{Phys. Rev. B} \textbf{\bibinfo{volume}{49}},
  \bibinfo{pages}{14251} (\bibinfo{year}{1994}).

\bibitem[{\citenamefont{Kresse and
  Furthm\"{u}ller}(1996{\natexlab{a}})}]{kressevasp3}
\bibinfo{author}{\bibfnamefont{G.}~\bibnamefont{Kresse}} \bibnamefont{and}
  \bibinfo{author}{\bibfnamefont{J.}~\bibnamefont{Furthm\"{u}ller}},
  \bibinfo{journal}{Comput. Mat. Sci.} \textbf{\bibinfo{volume}{6}},
  \bibinfo{pages}{15} (\bibinfo{year}{1996}{\natexlab{a}}).

\bibitem[{\citenamefont{Kresse and
  Furthm\"{u}ller}(1996{\natexlab{b}})}]{kressevasp4}
\bibinfo{author}{\bibfnamefont{G.}~\bibnamefont{Kresse}} \bibnamefont{and}
  \bibinfo{author}{\bibfnamefont{J.}~\bibnamefont{Furthm\"{u}ller}},
  \bibinfo{journal}{Phys. Rev. B} \textbf{\bibinfo{volume}{54}},
  \bibinfo{pages}{11169} (\bibinfo{year}{1996}{\natexlab{b}}).

\bibitem[{\citenamefont{G.~Kresse}(1999)}]{kresse}
\bibinfo{author}{\bibfnamefont{D.~J.} \bibnamefont{G.~Kresse}},
  \bibinfo{journal}{Phys. Rev. B} \textbf{\bibinfo{volume}{59}},
  \bibinfo{pages}{1758} (\bibinfo{year}{1999}).

\bibitem[{\citenamefont{Bl\"ochl}(1994)}]{bloechl}
\bibinfo{author}{\bibfnamefont{P.~E.} \bibnamefont{Bl\"ochl}},
  \bibinfo{journal}{Phys. Rev. B} \textbf{\bibinfo{volume}{50}},
  \bibinfo{pages}{17953} (\bibinfo{year}{1994}).

\bibitem[{\citenamefont{Perdew et~al.}(1996)\citenamefont{Perdew, Burke, and
  Ernzerhof}}]{perdew}
\bibinfo{author}{\bibfnamefont{J.~P.} \bibnamefont{Perdew}},
  \bibinfo{author}{\bibfnamefont{K.}~\bibnamefont{Burke}}, \bibnamefont{and}
  \bibinfo{author}{\bibfnamefont{M.}~\bibnamefont{Ernzerhof}},
  \bibinfo{journal}{Phys. Rev. Lett.} \textbf{\bibinfo{volume}{77}},
  \bibinfo{pages}{3865} (\bibinfo{year}{1996}).

\bibitem[{\citenamefont{Perdew et~al.}(1997)\citenamefont{Perdew, Burke, and
  Ernzerhof}}]{perdew2}
\bibinfo{author}{\bibfnamefont{J.~P.} \bibnamefont{Perdew}},
  \bibinfo{author}{\bibfnamefont{K.}~\bibnamefont{Burke}}, \bibnamefont{and}
  \bibinfo{author}{\bibfnamefont{M.}~\bibnamefont{Ernzerhof}},
  \bibinfo{journal}{Phys. Rev. Lett.} \textbf{\bibinfo{volume}{78}},
  \bibinfo{pages}{1396} (\bibinfo{year}{1997}).

\bibitem[{\citenamefont{Hohenberg and Kohn}(1964)}]{DFT}
\bibinfo{author}{\bibfnamefont{P.}~\bibnamefont{Hohenberg}} \bibnamefont{and}
  \bibinfo{author}{\bibfnamefont{W.}~\bibnamefont{Kohn}},
  \bibinfo{journal}{Phys. Rev.} \textbf{\bibinfo{volume}{136}},
  \bibinfo{pages}{B864} (\bibinfo{year}{1964}).

\bibitem[{\citenamefont{Kohn and Sham}(1965)}]{KS}
\bibinfo{author}{\bibfnamefont{W.}~\bibnamefont{Kohn}} \bibnamefont{and}
  \bibinfo{author}{\bibfnamefont{L.~J.} \bibnamefont{Sham}},
  \bibinfo{journal}{Phys. Rev.} \textbf{\bibinfo{volume}{140}},
  \bibinfo{pages}{A1133} (\bibinfo{year}{1965}).

\bibitem[{\citenamefont{Abrahamson}(1969)}]{born-mayer}
\bibinfo{author}{\bibfnamefont{A.~A.} \bibnamefont{Abrahamson}},
  \bibinfo{journal}{Phys. Rev.} \textbf{\bibinfo{volume}{178}},
  \bibinfo{pages}{76} (\bibinfo{year}{1969}).

\end{thebibliography}
\end{document}